\documentclass[graybox]{svmult}

\usepackage{mathptmx}       
\usepackage{helvet}         
\usepackage{courier}        
\usepackage{type1cm}        
\usepackage{makeidx}         
\usepackage{graphicx}        
\usepackage{multicol}        
\usepackage[bottom]{footmisc}

\usepackage{multirow}

\usepackage{algorithmicx}
\usepackage{algorithm}
\usepackage{algpseudocode}
\makeatletter
\def\BState{\State\hskip-\ALG@thistlm}
\makeatother

\makeindex             

\begin{document}

\title*{BoostNet: Bootstrapping detection of socialbots, \\ and a case study from Guatemala}
\author{E.I. Velazquez Richards$^{1}$, E. Gallagher, \and P. Su\'arez-Serrato$^{1,2}$ }
\institute{$^{1}$ Instituto de Matem\'aticas, Universidad Nacional Aut\'onoma de M\'exico, \\ Ciudad Universitaria, Coyoac\'an, 04510, Mexico City, M\'exico.\\$^{2}$ Department of Mathematics, University of California Santa Barbara, Goleta, CA, USA\\ \email{pablo@im.unam.mx}\\ \copyright\, The Authors 2019. \today}


\maketitle
\abstract{We present a method to reconstruct networks of socialbots given minimal input. Then we use Kernel Density Estimates of Botometer scores from 47,000 social networking accounts to find clusters of automated accounts, discovering over 5,000 socialbots. This statistical and data driven approach allows for inference of thresholds for socialbot detection, as illustrated in a case study we present from Guatemala.}

\section{Introduction}
\label{sec:1}
In this paper we analyze data from the social networking platform Twitter. We use a statistical approach, with bi-variate Kernel Density Estimates, to detect automated accounts (socialbots) at scale in a large dataset. We present our BoostNet algorithm, which allows for the detection of networks of socialbots in microblogs and social media platforms given a very small number of initial accounts. We illustrate its performance with empirical data collected from Twitter in relation to current events in Guatemala. 

To begin to describe some of the context of the events that have led to this particular social media situation, first we point out that the displacement of people due to armed conflict and corruption is a problem that affects many countries around the world. 
This phenomenon has strongly affected the Central American countries of Honduras and Guatemala. Nevertheless, currently the US enjoys the lowest level of  undocumented immigrants in US in a decade, according to a Pew Research Center analysis of government data \cite{Pew18}. The same study indicates that border apprehensions have declined for Mexicans but risen for other Central Americans.

What are the root causes of migration? Understanding these can better help prevent forced displacement of people and thus also the effects on societies that receive them. In a previous work we investigated the use of socialbots in Honduras in relation to protests alleging electoral fraud \cite{GSV19}.

Consider the case of Guatemala. The  International Commission against Impunity in Guatemala (CICIG https://www.cicig.org/ \cite{UN-CICIG})  was created in 2006 by the United Nations and Guatemala. It is an international body whose mission is to investigate and prosecute serious crime \cite{UN-CICIG}. 

An independent international body, CICIG investigates illegal security groups and clandestine security organizations in Guatemala. These are criminal groups believed to have infiltrated state institutions, fostering impunity and undermining democratic advances since the end of the armed conflict in the 1990s. The third impeachment against President Jimmy Morales for illicit electoral financing during his electoral campaign in 2015 was requested by the Attorney General and the CICIG.
\medskip

The mandate of the CICIG was set to end originally on September 3rd, 2019, but it has been cut abruptly short as Guatemalan President Morales ordered the CICIG to leave the country on January 7th, 2019 \cite{Linthicum2019}. 
\medskip

After we published our work on socialbots in Honduras \cite{GSV19} we were contacted by a Guatemalan journalist claiming that similar socialbots were acting against the population there. It was claimed that multiple Twitter accounts were being used to systematically intimidate and harass members of the CICIG and the media that covers their activities. In April 2018 we were provided with 19 seed accounts of potential socialbots that were notorious in this instance for their negative behaviour. 

\begin{figure}[ht!]
\sidecaption
\includegraphics[scale=.30]{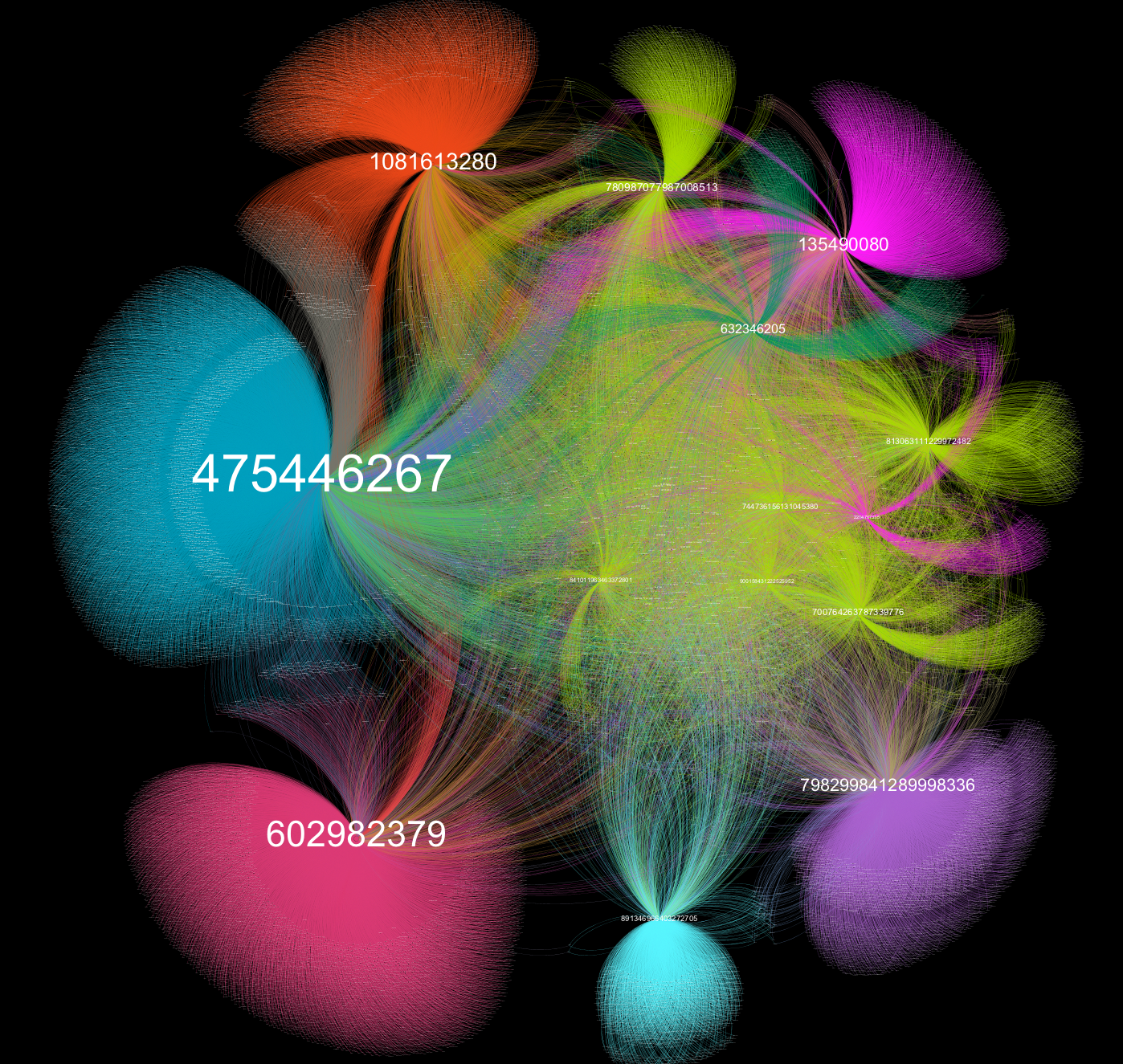}
\caption{ Gephi network graph created using OpenOrd and Force Atlas 2 force-directed layout algorithms. The network contains 35,208 nodes, 59,471 edges and 8 distinct clusters or communities. As per Twitter's data policies we have used the user ID to label the nodes, and not the account handle.} 
\label{fig:1}      
\end{figure}

\begin{figure}[ht!]
\sidecaption[t]
\includegraphics[scale=.30]{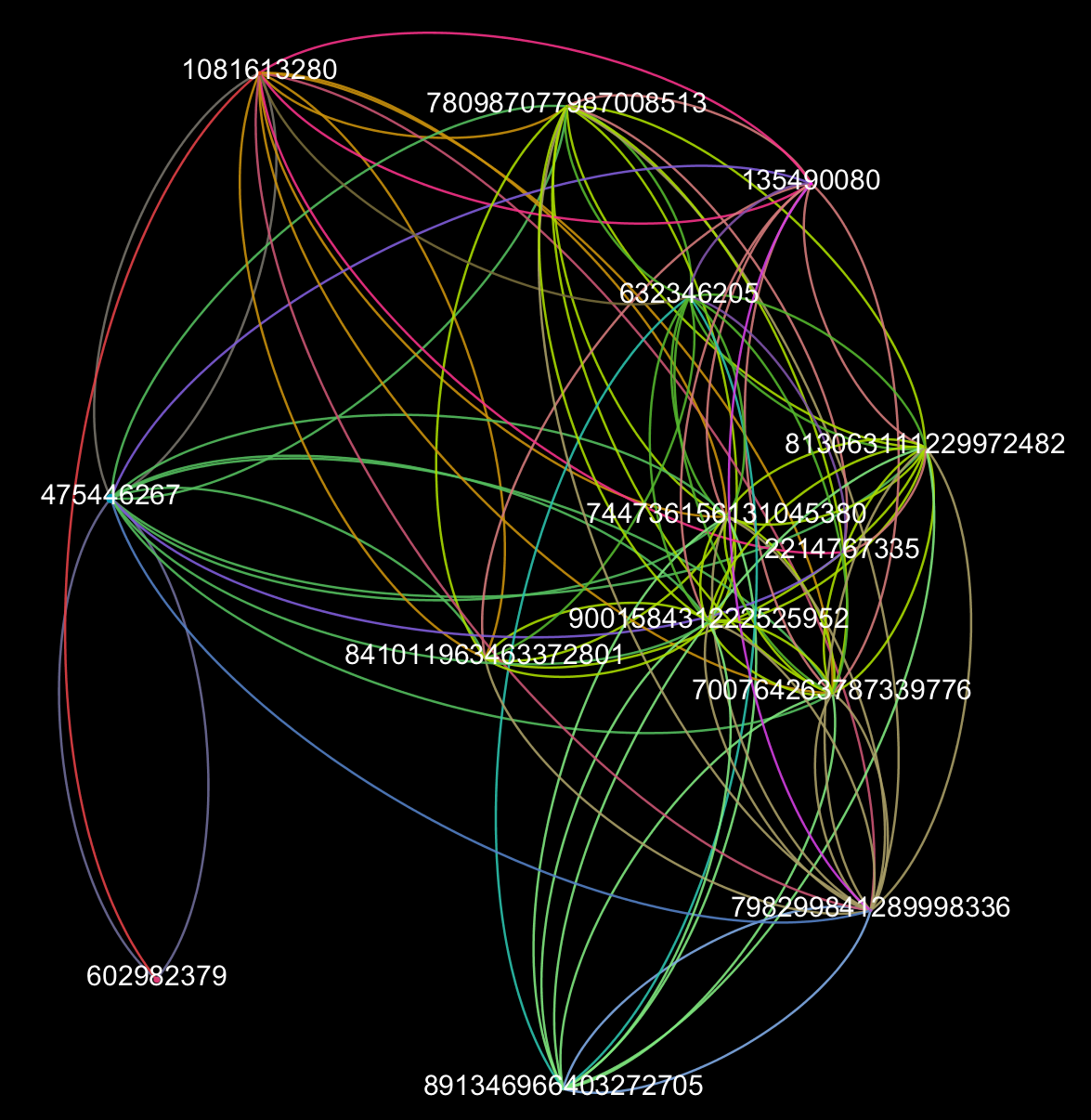}
\caption{Gephi network graph created using OpenOrd and Force Atlas 2 force-directed layout algorithms. The complete network (see Fig. 1) contains 35,208 nodes, 59,471 edges and 8 distinct clusters or communities, which was filtered by degree range 50 revealing 14 visible nodes (0.04 \%) and 100 visible edges (0.17 \%)of the complete network.}
\label{fig:2}      
\end{figure}

From these 19 accounts we reconstructed a network of over 35,000 accounts, by collecting their followers and their followees. 
 The rationale is that socialbot accounts are not generally followed by human accounts. Following this premise we begin with these 19 seed accounts and take two hops out into the follower network to find potential accounts that are also automated and being used for this purpose. This method, which we call BoostNet, is explained in Algorithm \ref{alg:algorithm} and the networks are visualized in figure \ref{fig:1} in terms of reach and spread of the full network, and in figure \ref{fig:2} in a subset of the most active bot accounts and their retweet relationships. This strategy led us to discover a socialbot network of over 3000 accounts. To this end we queried {\it Botometer} \cite{Davis2016} and performed a statistical analysis of the scores it provides to find the network of socialbots (explained below, see figure \ref{fig:botscores}).  We further validated our method by using 14 more accounts mentioned in a media interview about socialbot harassment in Guatemala from November 2018. From these 14 seed accounts we reconstructed a full network of over 12,000 accounts and found over 2,000 socialbots (see figure \ref{fig:botscores2}). There were over 600 socialbots common to both datasets. 

\medskip
In order to better understand the magnitude of this socialbot network, it is helpful to observe that Guatemala has a population of around 17 million people, and internet users include only 4.5 million \cite{Guate-Net}. Measurements of social media use in Guatemala indicate that 5.24\% of internet users are active on Twitter \cite{Guate-Tw}. We can therefore extrapolate an---admittedly rough---estimate of around 250,000 Twitter users in Guatemala (2018 figures). In this perspective, socialbot networks of 3,000 and 2,000 accounts can have a considerable impact.  

\medskip
{\bf Conclusions:}  
Our work here demonstrates how statistical methods can show the existence of considerable socialbot network of linked accounts. Given the potential size of Guatemala's total Twitter user base, the amount of socialbot accounts could certainly impede freedom of expression. These findings corroborate the experience of users (and journalists) who claimed wide-spread abuse of this technology for nefarious purposes was present in Guatemala.

Moreover, our BoostNet strategy can be employed in other circumstances and social media platforms, where limited observational data can then lead to a complete reconstruction of networks of malicious accounts. 

\section{Data Collection}
\label{sec:2}

In this section we describe our strategy to gather a large network of socialbot accounts from a small number of accounts that are reported to be abusing a social media service. We present an algorithm that can be replicated in other circumstances, and can be easily implemented to reconstruct a complete network of linked accounts. 

\subsection{ BoostNet: A method to find socialbot networks with minimal input}

The following pseudo-code illustrates our work-flow to construct networks where the human and socialbot accounts can be analyzed. Our method allows us to find large networks of socialbots given a small number of starting accounts. We illustrate its performance with an empirical case study here, we discovered two sets of socialbots; one containing over 3,000 socialbot accounts and the second containing over 2,000 socialbot accounts, starting from only 19 and 14 accounts respectively in each case that were reportedly harassing journalists and members of the CICIG. 

\begin{algorithm}
	\caption{ BoostNet : Bootstrapping Socialbot Network Detection}\label{alg:algorithm}
	\begin{algorithmic}[1]
	\Require A collection $C$ of Twitter Accounts
	\Ensure Full linked network $N(C)$ of with Socialbot account score \\
	{\bf Initialize}  \\
	For each account  $a$ in $C$:\\
	\hspace{2.1em} Collect followers $F(C)$ of the collection $C$ from Twitter's Rest API\\
	\hspace{2.1em} Collect those accounts  $FR(a)$ who are following $a$ from Twitter's Rest API \\
	\hspace{2.1em} Obtain scores of every account in $F(C)$, $FR(a)$ and $a$ from Botometer, to determine if it is Human or Socialbot \\
	 Construct a follower-followee network $N(C)$ annotated with Botometer scores\\
	 \Return $N(C)$\\
	 {\bf End}
	\end{algorithmic}
\end{algorithm}

\subsection{Comparison with Twitter's Stream API}

One poignant criticism of certain Twitter studies is the reliance on Twitter's Streaming API for data acquisition. While Twitter's Streaming API provides free and public access to a sample of tweets and has promoted research into social networks, there are certain limitations that its sampling method impose. Here we circumvent these difficulties in finding networks of linked accounts. Connections of followers and followees were queried from Twitter's Rest API. In this way we have reconstructed a full dataset of accounts that are linked in the same connected network.

Certain studies have avoided this sampling bias uncertainty from Twitter's Streaming API by using the Search API to obtain complete datasets \cite{Stella18}. Another option seems to be to work directly with Twitter, and some research has been successful at establishing influence relations using this kind of access \cite{Aral18}.

For this work we have reconstructed a full dataset of interest for our research using the Rest API only.

\section{Statistical Detection of Socialbot Networks}

For a review of {\em Botometer}, we recommend \cite{RiseSocialBots2016}. Socialbots have been  employed for political purposes \cite{Wooley2016}. It has also been observed that this technology is used in marketing and propaganda \cite{Varol2017-2}. Although research has uncovered other successful methods of bot detection  \cite{chavoshi2016identifying,Dickerson2014,Chu2012,Clark2015}, {\em Botometer} provides public API access.  The features it has built in, as well as a review of how it compares to, and surpasses, other methods can be found in \cite{Davis2016, Varol2017}. 

We have previously used this method for identifying bots in online communities in Latin America, specifically in Mexico and Honduras  \cite{suarez2016influence, GSV19, VYS18}. 

\begin{figure}[ht!]
\includegraphics[scale=.74]{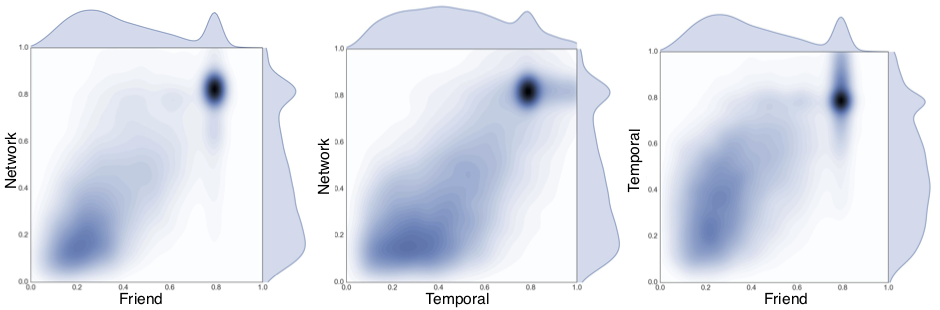}
\caption{2D Kernel decomposition estimate for Network-Friend, Network-Temporal, and Temporal-Friend, pairwise classifiers from Botometer, for the 35,308 Twitter accounts in our 1st dataset, obtained through our BoostNet method. The regions in the upper right corners correspond to the over 3,000 socialbot accounts that we discovered. These results were obtained on April 9th-18th 2018.}
\label{fig:botscores}       
\end{figure}

\begin{figure}[ht!]
\includegraphics[scale=.13]{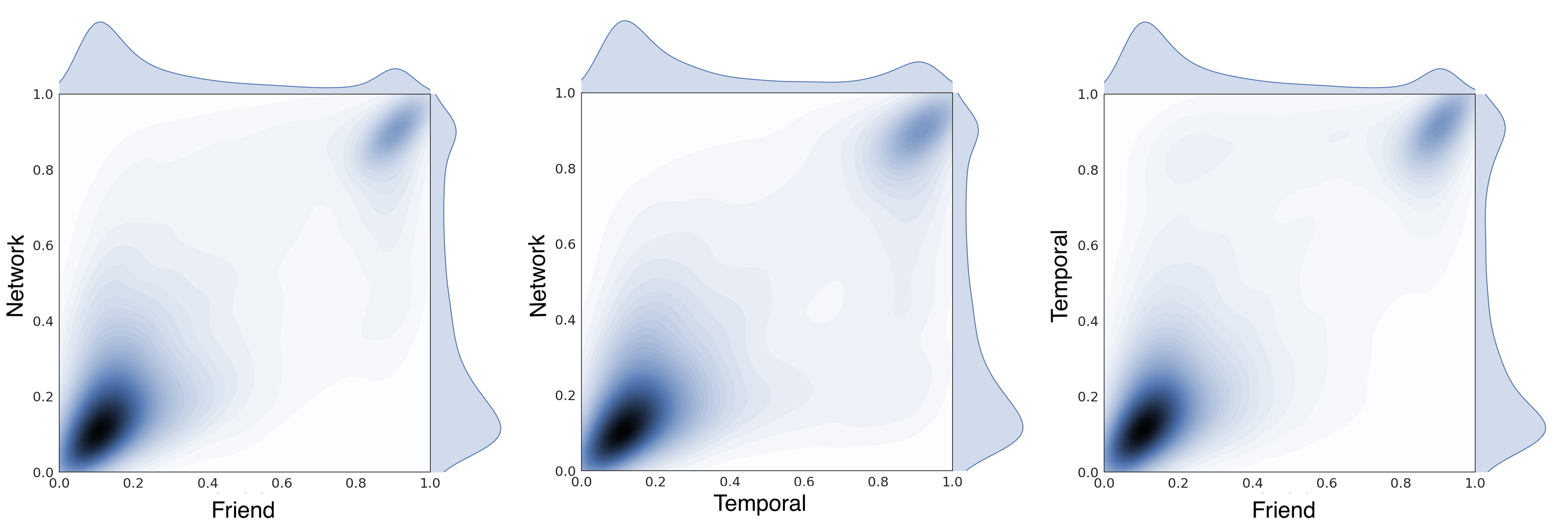}
\caption{2D Kernel decomposition estimate for Network-Friend, Network-Temporal, and Temporal-Friend, pairwise classifiers from Botometer, for the 12,044 Twitter accounts in our 2nd dataset, obtained through our BoostNet method. The regions in the upper right corners correspond to the over 2154 socialbot accounts that we discovered. These results were obtained between November 2018 and January 2019.}
\label{fig:botscores2}       
\end{figure}

In this work we have concentrated on three of the non-language specific classifiers that Botometer provides. Using the scores from Temporal, Network, and Friend evaluations that each account in our dataset yields, we aggregate this data and then find a 2D bimodal behaviour using KDE, as illustrated in figures \ref{fig:botscores} and \ref{fig:botscores2}. A numerical summary of the number of accounts found appears in table \ref{tab:accounts}.

\begin{table}[h]
\begin{center}	
\begin{tabular}{ccc c ccc}
 \multicolumn{6}{c}{Accounts in Network}\\ \cline{1-6}\\
	& &  &  Total & & Bots \\ \cline{4-6}
	&&&&&\\
	April &
	\multicolumn{1}{ c }{} & &  35208 & & 3009      \\
		& & & & & & \\
	November&
	\multicolumn{1}{ c }{} &  & 12044 &  & 2154 & \\ \cline{1-6}
\end{tabular} \\ \parbox{0.6\textwidth}{\caption{\label{tab:accounts}Linked accounts in our two datasets, from the full networks BoostNet reconstructed in April and in November 2018. A comparison between both datasets yields 3688 shared accounts, and at least 646 of them were classified as socialbots.}}
\end{center}
\end{table}

\begin{acknowledgement} We thank the OSoMe team in Indiana University for access to {\em Botometer }, and also Twitter for allowing access to data through their APIs. PSS acknowledges support from UNAM-DGAPA-PAPIIT-IN104819. 
\end{acknowledgement}

 \bibliographystyle{}
 \bibliography{}
%

\end{document}